\RequirePackage[mathlines]{lineno}
\documentclass[aps,prl,twocolumn,nofootinbib,showpacs,amsmath,amssymb]{revtex4-2}
\usepackage{overpic,graphicx}
\usepackage{dcolumn}
\usepackage{bm}
\usepackage{rotating}
\usepackage{subfigure}
\usepackage{color}
\usepackage[bookmarksnumbered, pdfstartview=FitH,colorlinks,urlcolor=blue, citecolor=blue,linkcolor=blue,] {hyperref}
\usepackage{lineno}
\usepackage{multirow}
\usepackage{amsmath}
\usepackage{makecell}

\newcommand{\MeasMass}{$M = 4188.8 \pm 4.7 \pm 8.0 \, {\rm MeV/} c^2$}
\newcommand{\MeasWidth}{$\Gamma =49 \pm 16 \pm 19 \,{\rm MeV}$}

\begin{document}


\title{\boldmath Measurement of  the Born cross section for $e^{+}e^{-}\to \eta h_c $  at center-of-mass energies between  4.1 and 4.6\,GeV}
\author{M.~Ablikim$^{1}$, M.~N.~Achasov$^{4,c}$, P.~Adlarson$^{75}$, O.~Afedulidis$^{3}$, X.~C.~Ai$^{80}$, R.~Aliberti$^{35}$, A.~Amoroso$^{74A,74C}$, Q.~An$^{71,58,a}$, Y.~Bai$^{57}$, O.~Bakina$^{36}$
% ...
\\
\vspace{0.2cm}
(BESIII Collaboration)\\
\vspace{0.2cm} {\it
$^{1}$ Institute of High Energy Physics, Beijing 100049, People's Republic of China\\
% ...
$^{3}$ Bochum  Ruhr-University, D-44780 Bochum, Germany\\
$^{4}$ Budker Institute of Nuclear Physics SB RAS (BINP), Novosibirsk 630090, Russia\\
% ...
$^{35}$ Johannes Gutenberg University of Mainz, Johann-Joachim-Becher-Weg 45, D-55099 Mainz, Germany\\
$^{36}$ Joint Institute for Nuclear Research, 141980 Dubna, Moscow region, Russia\\
% ...
$^{57}$ Southeast University, Nanjing 211100, People's Republic of China\\
$^{58}$ State Key Laboratory of Particle Detection and Electronics, Beijing 100049, Hefei 230026, People's Republic of China\\
% ...
$^{71}$ University of Science and Technology of China, Hefei 230026, People's Republic of China\\
% ...
$^{74}$ University of Turin and INFN, (A)University of Turin, I-10125, Turin, Italy; (B)University of Eastern Piedmont, I-15121, Alessandria, Italy; (C)INFN, I-10125, Turin, Italy\\
$^{75}$ Uppsala University, Box 516, SE-75120 Uppsala, Sweden\\
% ...
$^{80}$ Zhengzhou University, Zhengzhou 450001, People's Republic of China\\
\vspace{0.2cm}
$^{a}$ Deceased\\
% ...
$^{c}$ Also at the Novosibirsk State University, Novosibirsk, 630090, Russia\\
% ...
}
}
\begin{abstract}
We measure the Born cross section for the reaction $e^{+}e^{-}
\rightarrow \eta h_c$ from $\sqrt{s} = 4.129$ to $4.600$~GeV using
data sets collected by the BESIII detector running at the BEPCII
collider. A resonant structure in the cross section line shape near
4.200~GeV is observed with a statistical significance of 7$\sigma$.
The parameters of this resonance are measured to be \MeasMass\ and
\MeasWidth, where the first uncertainties are statistical and the
second systematic.
\end{abstract}
\pacs{13.66.Bc, 14.40.Gx}
\maketitle

\oddsidemargin  -0.2cm
\evensidemargin -0.2cm


In recent years, several new vector resonances, including the
$Y(4230)$, $Y(4360)$ and $Y(4660)$, have been observed in the
charmonium region at B-factories and $\tau$-charm
factories~\cite{BaBar:2005hhc,BaBar:2006ait,Belle:2007umv}.  While the potential
models~\cite{Barnes:2005pb} can accommodate the $\psi(4040)$,
$\psi(4160)$, and $\psi(4415)$, the new states appear to be
supernumerary.  Several models have been proposed to explain them as
exotic non-$c\bar{c}$
mesons~\cite{Close:2005iz,Ebert:2005nc,Liu:2005ay,Chen:2015dig,Bugg:2008wu}. The
nature of the first observed vector charmonium-like state, the
$Y(4230)$ (also known as $\psi(4230)$) is still mysterious.  It is
regarded as a good candidate for a hybrid state because its mass is
close to the vector hybrid state predicted by lattice
QCD~\cite{HadronSpectrum:2012gic} and also because of its small
electronic width~\cite{Chen:2016ejo,Ono:1984dp} and decay
pattern~\cite{Kou:2005gt}. But it is also interpreted as a
conventional
charmonium~\cite{Cao:2020vab,Li:2009zu,Llanes-Estrada:2005qvr},
hadronic molecule~\cite{Qin:2016spb,Close:2009ag}, non-resonant
enhancement~\cite{Coito:2019cts}, etc. More experimental measurements
are valuable to elucidate the nature of the $Y(4230)$.

Recently, BESIII performed a precise measurement of the cross section
of $e^+e^-\to\pi^+\pi^- J/\psi$ at center-of-mass (c.m.) energies from
3.774 to 4.600 GeV, in which two resonances were observed, namely the
$Y(4230)$ and the $Y(4320)$~\cite{BESIII:2016bnd}. This observation
has challenged our initial understanding of the $Y(4230)$.  A similar
double-resonance structure is also observed in other hadronic
modes, such as the $e^+e^-\to\pi^+\pi^-h_c$~\cite{BESIII:2016adj} and
$e^+e^-\to\pi^+\pi^-\psi(3686)$~\cite{BESIII:2017tqk} processes.
Studying the transitions from vector states to the $h_c$ meson is
particularly interesting because strong coupling to $h_{c}$ is
indicative of a hybrid state with a $c\bar{c}$ pair in a spin-singlet
configuration~\cite{Oncala:2017hop}.

The process $e^{+}e^{-} \to \eta h_c$ was previously observed for the
first time at $\sqrt{s}=$ 4.226 GeV by BESIII using data collected in
2012-2014~\cite{BESIII:2017dxi}. A hint of a resonance around
4.200~GeV was observed in the c.m. energy-dependent cross
section. Recently, the BESIII experiment collected more data at
c.m. energies from 4.129 to 4.600\,GeV. In this Letter we use the new
data sets to update the study of $e^{+}e^{-}\to\eta h_c$ with $h_c \to
\gamma \eta_c$ and $\eta \to \gamma \gamma$ to substantially improve
our measurement of the cross section. The total integrated luminosity
is measured to be $\rm \sim15\,fb^{-1}$ using large-angle Bhabha
scattering events with an uncertainty of
$1.0\%$~\cite{BESIII:2015qfd}, and the c.m.\ energies are measured
using the di-muon process~\cite{BESIII:2015zbz}.

The BESIII detector is described in detail
elsewhere~\cite{BESIII:2009fln}. The determination of the detection
efficiency and estimation of physics backgrounds are carried out with
Monte Carlo (MC) samples. {\sc
  Geant4}-based~\cite{GEANT4:2002zbu,Allison:2006ve} detector
simulation software is used to model the detector response.  The
signal MC events are simulated for each decay mode of the $\eta_c$
meson with {\sc kkmc}~\cite{Jadach:1999vf} and {\sc
  besevtgen}~\cite{Ping:2008zz}, in which the line shape of $\eta_c$
is a Breit-Wigner (BW) function.  The $e^{+}e^{-}\to\eta h_c$ process
is assumed to be dominated by the S wave, and the E1 transition
$h_c\to\gamma\eta_c$ is simulated using the dedicated helicity
formalism~\cite{Peters:2004qw}. In order to study potential
backgrounds, inclusive MC samples are simulated at each c.m. energy
with {\sc kkmc}.  These MC samples consist of charmed meson
production, initial state radiation (ISR) production of the low-mass
vector charmonium states, QED events and continuum processes.  The
known decay modes of the resonances are simulated with {\sc besevtgen}
with branching fractions set to the world average
values~\cite{ParticleDataGroup:2020ssz}, and the remaining events
associated with charmonium decays are simulated with {\sc
  lundcharm}~\cite{Chen:2000tv}. Other hadronic events are simulated
with {\sc pythia}~\cite{Sjostrand:2014zea}.

In this measurement, we first reconstruct the E1 photon and bachelor
$\eta$, and then the $\eta_c$ is reconstructed with sixteen hadronic final
states with a total branching fraction of about 40\%: $p\bar{p}$,
$2(\pi^+ \pi^-)$, $2(K^+ K^-)$, $K^+ K^- \pi^+ \pi^-$, $p\bar{p} \pi^+
\pi^-$, $3(\pi^+ \pi^-)$, $K^+ K^- 2(\pi^+ \pi^-)$, $K^+ K^- \pi^0$,
$p \bar{p}\pi^0$, $K^0_S K^\pm \pi^\mp$, $K^0_S K^\pm\pi^\mp \pi^\pm
\pi^\mp$, $\pi^+ \pi^- \eta$, $K^+ K^- \eta$, $2(\pi^+ \pi^-) \eta$,
$\pi^+ \pi^- \pi^0 \pi^0$, and $2(\pi^+\pi^-) \pi^0 \pi^0$. The $\eta
/\pi^{0}$ candidates are reconstructed using two photons and the
$K_S^0$ candidates are reconstructed via the $\pi^+\pi^-$ decay
channel. For the selected candidates, we apply a fit to the
distribution of the $\eta$ recoil mass to obtain the $\eta h_c$ signal
yield.

The selection criteria for charged tracks and photon candidates as
well as the $\pi^0$, $\eta$ and $K_S^0$ reconstruction are described
in Ref.~\cite{BESIII:2017dxi}. After this selection, a four-constraint
(4C) kinematic fit is performed for each event imposing overall
energy-momentum conservation, and the $\chi^2_{\rm 4C}$ is required to
be less than 25 to suppress background events.  The best candidates of
$\pi^0$, $\eta$ and $K_S^0$ as well as the particle identification
(PID) assignments of charged tracks in an event are determined by
minimizing $\chi^2 \equiv \chi^2_{\rm 4C}+\chi^2_{\rm
  1C}+\chi^2_{\rm PID}+\chi^2_{\rm vertex}$, where $\chi^2_{\rm 1C}$
is the overall $\chi^2$ of the 1C fit for all $\pi^0$ and $\eta$
candidates, $\chi^2_{\rm PID}$ is the sum over all charged tracks of
the $\chi^2$ of the PID hypotheses, and $\chi^2_{\rm vertex}$ is the
$\chi^2$ of the $K_S^0$ secondary-vertex fit. If there is no $\pi^0$
($K_S^0$) in an event, the corresponding $\chi^2_{\rm 1C}$
($\chi^2_{\rm vertex}$) is set to zero. If more than one $\eta$
candidate with recoil mass in the $h_c$ pre-selection signal region
[3.480, 3.600] GeV$/c^2$ is found, the one with mass of the $\eta_c$
candidate closest to $\eta_{c}$ known mass is selected.

The requirements on $\chi_{\rm4C}^{2}$ and the mass windows for $\eta$
and $\eta_{c}$ selection are determined by maximizing the
figure-of-merit, which is defined as $S/\sqrt{S+B}$. Here, $S$ and $B$
are the signal and background yields, respectively.  The optimization
was performed using the combined statistics of four samples with high
integrated luminosity taken at $\sqrt{s}$ = 4.179, 4.189, 4.199 and
4.209 GeV.

\begin{figure}[htbp]
	\begin{center}		
		\begin{overpic}[width=1.0\columnwidth,angle=0]{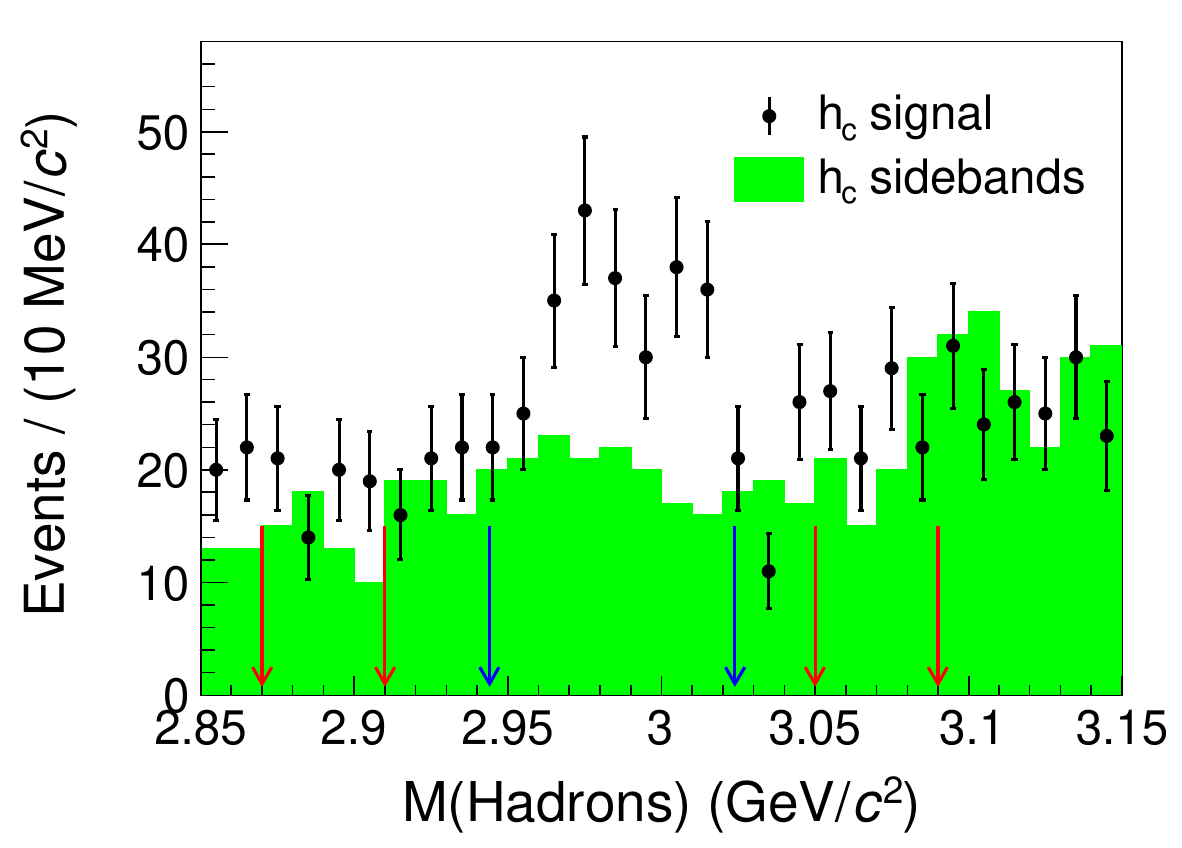}
		\end{overpic}
	\end{center}
	\caption{ Combined invariant mass distribution of the sixteen
          hadronic final states forming the $\eta_c$ meson in the
          $h_c$ signal (dots with error bar) and sideband ranges
          (green shaded histogram) at $\sqrt{s}$ = 4.179\,GeV. The
          blue arrows show $\eta_c$ signal region and the red arrows
          indicate $\eta_c$ sidebands. }\label{hc_etac_bg}
	
\end{figure}

\begin{figure}[tbhp]
	\begin{center}
		\begin{overpic}[width=1.0\columnwidth,angle=0]{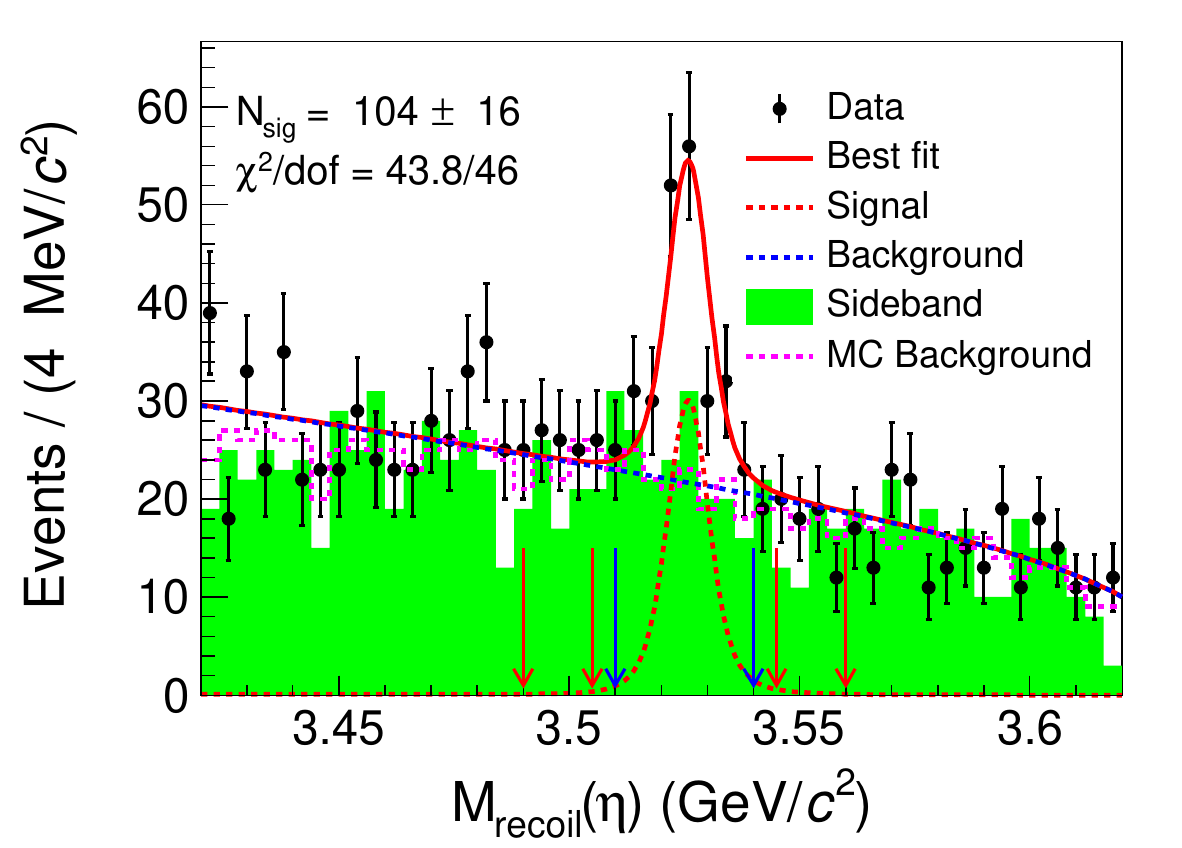}
		\end{overpic}
		\caption{Sum of the simultaneous fits to the $\eta$
                  recoil mass spectra for all sixteen $\eta_c$ decay
                  modes at $\sqrt{s}=4.179$\,GeV. The dots with error
                  bars represent the $\eta$ recoil mass spectrum in
                  data. The solid red line shows the total fit
                  function, and the dashed red and blue lines are the
                  signal and background components of the fit. The
                  green shaded histogram shows the events from
                  $\eta_c$ sidebands. The dashed pink line is the
                  simulated background. The blue arrows show $h_c$
                  signal region and the red arrows indicate $h_c$
                  sidebands.
			\label{fig:fitRMetasimtot}}
	\end{center}
\end{figure}

After applying all the above criteria and using a $h_c$ mass
window of [3.510, 3.540]\,${\rm GeV}/c^2$, a clear $\eta_c$ signal is
observed in the invariant mass spectrum of hadrons in the data sample
taken at $\sqrt{s}=$ 4.179 GeV, which is shown in
Fig.~\ref{hc_etac_bg}.  A clear $h_{c}$ signal is also seen in the
$\eta$ recoil mass distribution when applying a mass window of [2.944,
  3.024]\,${\rm GeV}/c^2$ to the invariant mass distribution of
hadrons from $\eta_c$ decay, as shown in Fig.~\ref{fig:fitRMetasimtot}.

In the $\eta$ recoil mass spectra, as depicted in Fig.~\ref{fig:fitRMetasimtot}, 
no peaking structures are observed within the $\eta_{c}$ sidebands, 
which are delineated as [2.870, 2.910]\,${\rm GeV}/c^2$ and [3.050, 3.090]\,${\rm GeV}/c^2$, 
as demonstrated in Fig.~\ref{hc_etac_bg}.
Likewise, the hadronic invariant mass spectrum resulting from $\eta_c$ decay  within the $h_c$ sidebands, 
specified as [3.490, 3.505]\,${\rm GeV}/c^2$ and [3.545, 3.560]\,${\rm GeV}/c^2$, exhibits no peak formations, 
as illustrated in Fig.~\ref{hc_etac_bg}.  
In addition, inclusive MC samples simulated at $\sqrt{s}=4.179\,\rm{GeV}$ are
analyzed with the same event selection as applied to data, and the
dominant background is found to be the continuum processes. Other
sources just contribute negligible background. The comparison
between data and the inclusive MC sample is shown in
Fig.~\ref{fig:fitRMetasimtot}.

To obtain the $h_c$ yield, the
sixteen $\eta$ recoil mass distributions are fitted simultaneously
with an unbinned maximum likelihood method. In the fit, the signal
shape is determined by MC simulation, and the background is described
by an ARGUS function~\cite{ARGUS:1990hfq}.  The truncation point is
set to be the same for all channels and fixed according to simulation
at each c.m. energy. The total signal yield, $N_{sig}$, is the sum of
the yields from all the sixteen $\eta_c$ decay channels. The signal
yield of the $i$-th channel is set to be $N_{\rm sig}\cdot f_i$, where
$f_i$ is the weight factor $f_i=\epsilon_{i}\mathcal{B}_{i} /\sum
\epsilon_{i} \mathcal{B}_i$, $\mathcal{B}_{i}$ denotes the branching
fraction of $\eta_c$ decays to the $i$-th final state and
$\epsilon_{i}$ is the corresponding efficiency.  The fraction of the
background in each mode is free.  The sum of the fit results
at $\sqrt{s}=4.179$\,GeV is shown in
Fig.~\ref{fig:fitRMetasimtot}. The corresponding $\chi^2$ per degree
of freedom (dof) for this fit is $\chi^2/\rm{dof}=43.8/46=0.95$. The
total signal yield is $104\pm16$ with a statistical significance of
$9.6\sigma$.
Using the same method, we also study the data samples taken at other
c.m. energies. Significant signals (more than $5\sigma$) are observed
at $\sqrt{s}=$ 4.179, 4.189 and 4.226 GeV, and evidence (between
$3\sigma$ and $5\sigma$) is found at $\sqrt{s}=$ 4.209, 4.358 and
4.436 GeV.

The Born and ``Dressed" cross sections are calculated by the following
formula:
\begin{equation}\label{eq1}
	\begin{aligned}
		\sigma^{\rm Born}&=\frac{\sigma^{\rm Dressed}}{|1+\Pi|^2} \\
		&=\frac{N_{\rm sig}}{\mathcal{L} (1+\delta) |1+\Pi|^2 \mathcal{B}(\eta\to\gamma \gamma) \mathcal{B}({h}_c\to\gamma \eta_c)  \Sigma_{i}\epsilon_{i} \mathcal{B}_i},
	\end{aligned}
\end{equation}
where $\mathcal{L}$ is the integrated luminosity of the data sample
taken at each c.m. energy.  The radiative correction factor
$(1+\delta)$ at each c.m. energy is calculated
iteratively~\cite{BESIII:2017dxi}.  The term $|1+\Pi|^2$ is the
vacuum-polarization (VP) correction factor and is calculated according
to Ref.~\cite{Eidelman:1995ny}.  In addition,
$\mathcal{B}(\eta\to\gamma \gamma)$ and $\mathcal{B}(h_c\to\gamma
\eta_c)$ are the branching fractions of the $\eta$ and $h_c$ decays,
respectively, and $\epsilon_i$ is the efficiency determined with MC
simulation. The measured Born cross sections and the quantities that
enter the calculation at all c.m energies are listed in the
Supplemental Material~\cite{sup}.

The cross section of $e^+e^-\to \eta h_c$ from $\sqrt{s}=$ 4.129 to
4.600 GeV is parameterized as:
\begin{equation}
	\sigma^{\rm Dressed}(s) = \\ |{\rm BW}_{1}(s)+{\rm BW}_{2}(s)e^{i\phi}|^2+|{\rm BW}_{3}(s)|^2.
\end{equation}
The relativistic BW amplitude for a resonance $Y\to\eta h_c$ in the
fit is written as:
\begin{equation}
    {\rm BW}(s)=\frac{\sqrt{12\pi\Gamma_{ee}\Gamma_{\rm tot}{\cal B}(Y\to\eta h_c)}}{s-M^2+iM\Gamma_{\rm tot}}\sqrt{\frac{{\rm PS}(\sqrt{s})}{{\rm PS}(M)}},
\end{equation}
where $M$, $\Gamma_{\rm tot}$, $\Gamma_{ee}$, and ${\cal B}(Y\to\eta
h_c)$ are the mass, full width, electronic partial width , and
branching fraction of the corresponding resonance, respectively, and
$\sqrt{{\rm PS}(\sqrt{s})/{\rm PS}(M)}$ is the two-body phase space
factor~\cite{ParticleDataGroup:2020ssz}. It is important to note that
the definition of $\Gamma_{ee}$ includes vacuum polarization
effects. Consequently, the dressed cross sections are fitted rather
than the Born cross sections. A maximum likelihood fit is used to
obtain the parameters of these three resonances.  In the fit, the
parameters of the second BW function are fixed to that of the
$Y(4360)$ due to the large uncertainty of the cross section in this
region, while the other two BW functions are free.  Our analysis is
primarily concerned with the measurement of ${\rm
  BW}_1$. Consequently, the interference effects between ${\rm BW}_1$
and ${\rm BW}_3$, as well as between ${\rm BW}_2$ and ${\rm BW}_3$,
have been neglected in the nominal fit because of statistics for ${\rm
  BW}_2$ and ${\rm BW}_3$. There could be multiple solutions in the
fit due to interference, but we vary the initial values of $\phi$ and
only find one solution.

\begin{figure}[htbp]
	\begin{center}
		\begin{overpic}[width=1.00\columnwidth,angle=0]{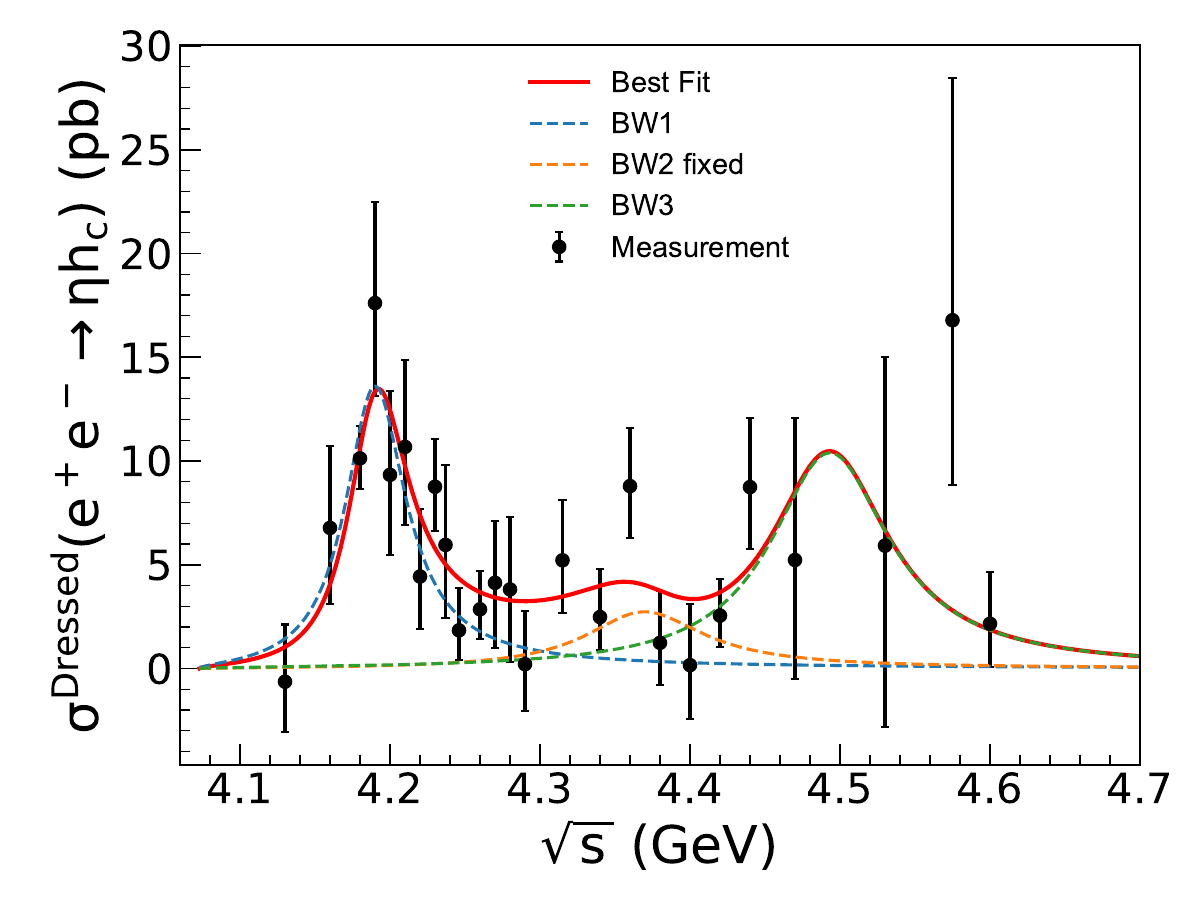}
			\put(20,55){(a)}
		\end{overpic}
		\begin{overpic}[width=1.00\columnwidth,angle=0]{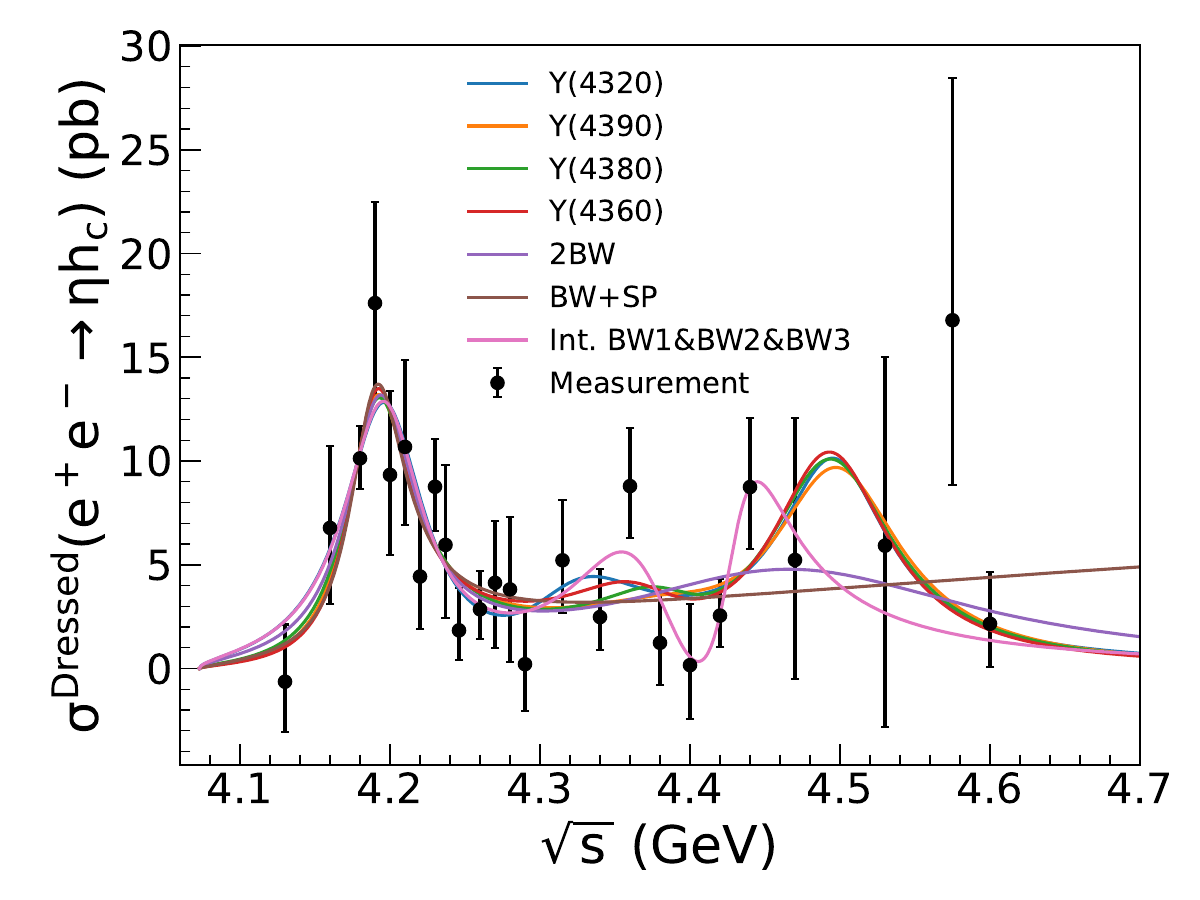}
			\put(20,55){(b)}
		\end{overpic}
	\end{center}
	\caption{(a) Result of the fit to the $\sqrt{s}$-dependent
          cross section $\sigma^{\rm Dressed}(e^+e^-\to \eta h_c)$.
          Dots with error bars are data, and the red solid curve shows
          the fit with three resonances. The dashed curves show the
          three individual resonances. (b) The comparison of the
          different models. The curves labeled as ``$Y(4320)$",
          ``$Y(4390)$", ``$Y(4380)$" and ``$Y(4360)$" show the models
          in which the parameters of ${\rm BW}_2(s)$ are fixed to
          $Y(4320)$, $Y(4390)$, $Y(4380)$ and $Y(4360)$,
          respectively. The curve ``2BW" represents the model without
          ${\rm BW}_2(s)$. The curve ``BW+SP" shows the model that sum
          of a BW function and phase space shape. The curve
          ``Int. BW1\&BW2\&BW3" shows the model that take the interference
          between three ${\rm BW}(s)$s into
          account.}\label{FitBESNEW}
\end{figure}

The fit results are shown in Fig.~\ref{FitBESNEW}(a) and the fitted
parameters of the BW functions are listed in Table~\ref{tab:Fit_para},
where the uncertainties are statistical only. The statistical
significance of the first resonance is calculated to be $8\sigma$,
which is obtained by comparing the change of the log-likelihood value
$\Delta(-{\rm ln}L) = 41$ and degrees of freedom $\Delta {\rm dof} =
4$ with and without this resonance in the fit.

To check the fit stability of the first resonance, we also use many
different parameterizations, including changing the
second resonant parameters to different hypotheses:
$Y(4320)$~\cite{BESIII:2016bnd}, $Y(4380)$~\cite{BESIII:2017tqk},
$Y(4390)$~\cite{BESIII:2016adj} and removing the second resonance in
the model.  In addition, we also use the sum of a BW function and
phase space shape to fit the cross section line shape and use a model
that takes the interference between all three resonances
into account.  The comparison of the fitted line shapes of
c.m. energy-dependent cross sections is shown in
Fig.~\ref{FitBESNEW}(b).  The choice of the fit model leads to the
dominant systematic uncertainties on the mass and width
parameters. The significance of the first resonance remains above
7$\sigma$ in all the alternative fits.

\begin{table}[htbp]
	\centering
	\caption{Results of the fit to the distribution of
          $\sigma^{\rm Dressed}(e^+e^-\to\eta h_c)$ for the first
          resonance. Here, $M$ and $\Gamma_{\rm tot}$ are the
          mass and total width of the resonance. $\Gamma_{ee}\mathcal{B}$
          is the product of the $e^+e^-$ partial width and branching
          fraction of $Y\to\eta h_c$. The first uncertainties are statistical and the
          second systematic.}
	\label{tab:Fit_para}
	\begin{tabular}{cccc}		
		\hline\hline
		  $\Gamma_{ee}\mathcal{B}$ (eV) & $M$ (MeV/$c^2$) & $\Gamma_{\rm tot}$ (MeV)    \\
		\hline		
		  $0.80 \pm 0.19 \pm 0.45$  & $4188.8 \pm 4.7 \pm 8.0$ & $49 \pm 16 \pm 19$ \\
		\hline
		\hline
	\end{tabular}
\end{table}

The systematic uncertainties for the measured Born cross sections are
determined as follows. The integrated luminosity is measured using
Bhabha events, with an uncertainty of $1.0\%$~\cite{BESIII:2015qfd}.
To estimate the uncertainty due to the data/MC mass resolution
difference in the fit to the recoil mass of $\eta$, the simulated
signal shape is shifted and convolved with a Gaussian
function to match the shape in data. The parameters of the Gaussian
function are obtained by a control sample of $e^+e^-\to\eta
J/\psi$. The average change of the cross section with and without this
correction among all the data sets is taken as the common systematic
uncertainty. To estimate the uncertainty due to the background shape,
a second order Chebyshev function instead of the ARGUS function is
used as an alternative model.  The average fit result difference
between these two background shapes in all data sets is adopted as the
systematic uncertainty. The systematic uncertainty from the fit range
is determined by changing the fit range randomly and redoing the
fit, and the average difference from the nominal result among all data sets
is taken as the systematic uncertainty. The branching fraction of
$h_c\to\gamma\eta_c$ is taken from Ref.~\cite{BESIII:2010gid}, and
its uncertainty, which is 15.7\%, will
propagate to the cross section measurement.  The ISR correction factor
at a given c.m. energy is determined using the cross section line
shape from the threshold to the c.m. energy of interest. The nominal
energy-dependent cross section is parameterized with the sum of three
BW functions as shown in Fig.~\ref{FitBESNEW}(a). The uncertainty of
the line shape is estimated by using different models as
discussed in the fit to the c.m. energy-dependent cross section (shown
in Fig.~\ref{FitBESNEW}(b)). The line shape of the cross section also
affects the efficiency. To consider this effect, we use the method
introduced in Ref.~\cite{BESIII:2017dxi}. To investigate the
uncertainty due to the vacuum polarization factor, we use two
available VP
parameterization~\cite{Eidelman:1995ny,WorkingGrouponRadiativeCorrections:2010bjp}. The
difference between them is 0.3\% and is taken as the systematic
uncertainty. In the simultaneous fit to the $\eta$ recoil mass
distributions, $\epsilon_i \mathcal{B}_i$ is used to constrain the
weights between different $\eta_c$ decay modes, so the uncertainty
from $\epsilon_i \mathcal{B}_i$ will affect the signal yield.  To
consider this issue, we use the refitting method introduced in
Ref.~\cite{BESIII:2017dxi} to estimate the uncertainty due to
$\epsilon_{i} \mathcal{B}_i$.  The systematic uncertainties related to
efficiency including charged track, photon, $K^0_{S}$, $\pi^0$ and
$\eta$ reconstruction, PID, kinematic fit, $\eta_c$ tag and cross feed
are estimated with the same method as described in
Ref.~\cite{BESIII:2017dxi}. The angular momentum between the $\eta$
and $h_c$ mesons is investigated from data and is found to be between
S and D waves. The uncertainty is estimated by comparing the
efficiencies from the pure S wave MC and mixture of S and D wave
MC. The uncertainties related to efficiency at $\sqrt{s}=4.179$ GeV
are given in the Supplemental Material~\cite{sup}.

\begin{table}[htbp]
	\centering
	\caption{Relative systematic uncertainties on $\sigma^{\rm Born}(e^+e^-\to \eta h_c)$ (in \%) at $\sqrt{s}=4.179$\,GeV.}
	\begin{tabular}{lc}
		\hline
		\hline
		Source    \hspace{5em}                                        & Uncertainty of $\sigma^{\rm Born}$  \\
		\hline
		Luminosity                                                             & $1.0$                         \\
		Signal shape                                                         & $1.4$                         \\
		Background shape                                                & $7.0$                         \\
		Fitting range                                                          & $3.6$                         \\
		$\mathcal{B}({h}_c\to\gamma \eta_c)\mathcal{B}(\eta\to\gamma \gamma)$                            & $15.7$                         \\
		ISR correction                                                       & $1.9$                         \\
		VP correction                                                        & $0.3$                         \\
		$\Sigma_{i}\epsilon_{i} \mathcal{B}_i$                       & $10.4$                         \\
		\hline
		Total                                                                     & $20.6$                          \\ 
		\hline
		\hline
	\end{tabular}
	\label{sys_summary}
\end{table}

The systematic uncertainties from different sources are listed in
Table~\ref{sys_summary}. All sources are treated as uncorrelated, so
the total systematic uncertainty is obtained by summing them in
quadrature.  For the data sets without significant $\eta h_c$ signals,
an upper limit of the cross section at the 90\% confidence level is
obtained using a Bayesian method. The systematic uncertainties are
taken into account by convolving the probability density function of
the measured cross section with a Gaussian
function~\cite{Stenson:2006gwf}.

The systematic uncertainties for the fit parameters to the cross
section line shape are described as follows. To estimate the
uncertainty for the fit model, we use all the variations described in
the fit to the c.m. energy-dependent cross section as shown in
Fig.~\ref{FitBESNEW}(b).  The largest deviation is taken as the
systematic uncertainty. The c.m. energies of all data sets are
measured using di-muon events with uncertainty $\pm1$ MeV.  The
uncertainty of the c.m. energy measurement will propagate to the mass
of the resonances directly.  The uncertainty due to the beam energy
spread is found to be negligible.  The systematic uncertainty for the
cross section measurements can be classified to two categories: the
uncertainties due to the fit, which are data-set independent and the
remaining uncertainties, which are treated as correlated among all
data-sets.  The first kind are studied by changing
the configuration in the fit to the $\eta$ recoil mass. We vary the
signal shape, background shape and fit range for each data set as
discussed before and then redo the fit to these alternatively
measured cross sections, and the largest deviations of the fitted
parameters of ${\rm BW}_1(s)$ are taken as the systematic
uncertainties. The second kind of uncertainties would not affect mass
and width of each resonance, but will propagate to the
$\Gamma_{ee}^{Y}\mathcal{B}$ by the same amount directly. The average
of the correlated systematic uncertainty for
$\Gamma_{ee}^{Y}\mathcal{B}$ is 19\%.
Table~\ref{tab:Res_Sys_Sum} summarizes the uncertainties of parameters
for the resonance around $\sqrt{s}=$ 4.200 GeV from the
c.m. energy-dependent cross sections.

\begin{table*}[htbp]
	\caption{The systematic uncertainties due to the fit of the
	cross section line shape. $\sqrt{s}$ refers to
	c.m. energy, and $\sigma^{\rm Dressed}_{\rm ind}$ and $\sigma^{\rm
	Dressed}_{\rm cor}$ represent the data-set independent and
	correlated uncertainties in the cross section measurements.}
	\label{tab:Res_Sys_Sum}
	\begin{center}
			\begin{tabular}{lcccccc}
				\hline\hline
				&  $\sqrt{s}$   &  Beam Spread  & Model  & $\sigma^{\rm Dressed}_{\rm ind}$ &  $\sigma^{\rm Dressed}_{\rm cor}$   &Sum   \\
				\hline
				$ M_1$ (MeV/$c^2$) & $1.0$ & $0.0$ & $7.6$ & $2.2$ & $\cdots$ & $8.0$ \\
				$\Gamma_{\rm tot}^{Y_1}$ (MeV)  & $\cdots$ & $0.1$ & $17.3$ & $8.2$ & $\cdots$ & $19.2$  \\
				$\Gamma_{ee}^{Y_1}\mathcal{B} $ (eV) & $\cdots$ & $0.0$ & $0.4$ & $0.1$ & $0.2$ & $0.5$  \\
				\hline
				\hline
			\end{tabular}
	\end{center}
\end{table*}

In summary, the $e^{+}e^{-}\to\eta h_c$ process is studied with
the data samples taken at c.m. energies from 4.129 to 4.600 GeV.  The
corresponding Born cross sections or upper limits of Born cross
sections at each c.m. energy are obtained. In the cross section
lineshape, a resonant structure near 4.200 GeV is observed with a
statistical significance of 7$\sigma$.  The parameters of this
resonance are measured to be: \MeasMass ~and \MeasWidth. Here, the
first uncertainties are statistical and second systematic. This result
is consistent with the parameters of $\psi(4160)$ but also not far
away from the $\psi(4230)$ observed from the $\pi^+\pi^- J/\psi$
process. The $1^{--}$ hybrid charmonium state predicted by the BOEFT
model~\cite{Berwein:2015vca} has a mass of $4.15\pm0.15$\,GeV,
therefore, our measurement is also consistent with this prediction.

\acknowledgments

The BESIII Collaboration thanks the staff of BEPCII and the IHEP computing center for their strong support. This work is supported in part by National Key R\&D Program of China under Contracts Nos. 2023YFA1606704, 2020YFA0406300, 2020YFA0406400; National Natural Science Foundation of China (NSFC) under Contracts Nos. 11635010, 11735014, 11835012, 11935015, 11935016, 11935018, 11961141012, 12025502, 12035009, 12035013, 12061131003, 12192260, 12192261, 12192262, 12192263, 12192264, 12192265, 12221005, 12225509, 12235017; the Chinese Academy of Sciences (CAS) Large-Scale Scientific Facility Program; the CAS Center for Excellence in Particle Physics (CCEPP); Joint Large-Scale Scientific Facility Funds of the NSFC and CAS under Contract No. U1832207; CAS Key Research Program of Frontier Sciences under Contracts Nos. QYZDJ-SSW-SLH003, QYZDJ-SSW-SLH040; 100 Talents Program of CAS; The Institute of Nuclear and Particle Physics (INPAC) and Shanghai Key Laboratory for Particle Physics and Cosmology; European Union's Horizon 2020 research and innovation programme under Marie Sklodowska-Curie grant agreement under Contract No. 894790; German Research Foundation DFG under Contracts Nos. 455635585, Collaborative Research Center CRC 1044, FOR5327, GRK 2149; Istituto Nazionale di Fisica Nucleare, Italy; Ministry of Development of Turkey under Contract No. DPT2006K-120470; National Research Foundation of Korea under Contract No. NRF-2022R1A2C1092335; National Science and Technology fund of Mongolia; National Science Research and Innovation Fund (NSRF) via the Program Management Unit for Human Resources \& Institutional Development, Research and Innovation of Thailand under Contract No. B16F640076; Polish National Science Centre under Contract No. 2019/35/O/ST2/02907; The Swedish Research Council; U. S. Department of Energy under Contract No. DE-FG02-05ER41374.

\bibliography{bib_eta_hc}{}

\end{document}